\documentclass[aps,pra,twocolumn,groupedaddress,longbibliography]{revtex4-1}
\usepackage{graphicx}
\usepackage{caption}
\DeclareCaptionLabelSeparator{dot}{. }
\makeatletter
\def\justified{
	\let\\\@normalcr
	\@rightskip\z@skip \rightskip\@rightskip
	\leftskip\z@skip
	\parindent 0em\relax
	\setlength{\parfillskip}{0pt plus 1fil}}
\DeclareCaptionJustification{justified}{\justified}
\usepackage{subcaption}
\usepackage{amsmath}
\usepackage{amssymb}
\usepackage{mathrsfs} 
\usepackage{braket}
\usepackage{units}
\usepackage{ragged2e}
\usepackage[colorlinks,urlcolor=blue ,citecolor=blue ,linkcolor=blue ]{hyperref}
\usepackage{xcolor}
\usepackage{nicefrac} 
\usepackage{lipsum}
\usepackage{nameref}
\usepackage{hyperref}
\usepackage{textcomp} 
\usepackage{urwchancal}
\usepackage{xcolor}
\usepackage{float}
\definecolor{darkgreen}{rgb}{0,0.5,0}
\usepackage{multirow}
\usepackage{tabularx}
\usepackage[normalem]{ulem}

\newcommand{\bs}{\boldsymbol}

\newcommand{\as}{\ensuremath{a_{\rm s}}}
\newcommand{\add}{\ensuremath{a_{\rm dd}}}
\newcommand{\edd}{\ensuremath{\epsilon_{\rm dd}}}
\newcommand{\km}{\ensuremath{k_{\rm m}}}
\newcommand{\kmax}{\ensuremath{k_{\rm max}}}
\newcommand{\kmin}{\ensuremath{k_{\rm min}}}
\newcommand{\qmax}{\ensuremath{q_{\rm max}}}
\newcommand{\qmin}{\ensuremath{q_{\rm min}}}
\newcommand{\nk}{\ensuremath{\tilde{n}^{(1)}}}
\newcommand{\alphac}{\ensuremath{\alpha_{\rm crit}}}
\newcommand{\noc}{\ensuremath{n_{0_{\rm c}}}}

\newcommand{\vk}{\ensuremath{\bs{k}}}
\newcommand{\vq}{{\bs q}}
\newcommand{\br}{\ensuremath{\bs{r}}}
\newcommand{\bR}{\ensuremath{\bs{R}}}

\newcommand{\erfc}{\ensuremath{\textrm{erfc}}}

\newcommand{\um}{\mu{\rm m}}

\newcommand{\Uint}{U} 

\newcommand{\TF}{\mathcal{F}}

\newcolumntype{Y}{>{\centering\arraybackslash}X}

\begin{document}
	
    \title{Effect of trap imperfections on the density of a quasi-two-dimensional uniform dipolar quantum Bose gas}
	\author{Thibault Bourgeois$^{1,2}$,  Lauriane Chomaz$^{1,\star}$.}
	
	\affiliation{%
		$^{1}$ Physikalisches Institut, Universität Heidelberg, Im Neuenheimer Feld 226, 69120, Heidelberg, Germany.\\%
        $^{2}$ Département de Physique, Ecole Normale Supérieure, 24 rue Lhomond, 75005, Paris, France.
	}	

	\date{\today}
	
\begin{abstract}
We theoretically investigate the impact of weak static perturbations of a flat potential on the density of a quasi-two-dimensional dipolar Bose gas. {We consider the perturbative effects of potential perturbations at first order and restrict to the mean-field stable regime. We first study cosinusoidal potential perturbations at a given spatial frequency; this allows us to understand the effects of optical lattices as well as of isolated momentum contributions in the Fourier decomposition of an arbitrary potential. We then study potential perturbations characterized by a static white-noise spectrum over a given momentum range; this captures the effects of inherent optical aberrations in setups that create uniform optical dipole traps.} 
\end{abstract}

\maketitle

\section{Introduction}

The achievement of quantum degeneracy in gases of particles with large dipole moments has opened new avenues of research in which long-range anisotropic dipole-dipole interactions (DDIs) play a crucial role~\cite{Chomaz2022dpa,Langen2024}. In such gases, the competition between DDIs, contact interactions and external trap geometry yields exotic behaviors. For example, novel many-body quantum phases have been found, including liquid-like droplets, droplet crystals, and supersolids~\cite{Chomaz2022dpa}. 

While experiments so far have relied on standard harmonic traps created by Gaussian beams, traps of different geometries may allow further exotic behaviors and new phases to emerge~\cite{Lu2010sdo,Rocuzzo2022sea,Hertkorn2021pfi,Zhang2021pos}.  Recent developments have enabled the creation of such exotic geometries in ultracold-gas experiments with alkali atoms~\cite{Atomtronic2021,Navon2021,Gauthier2021dhr} and are to be extended to dipolar particles, see also ref.~\cite{Juhasz2022htr}. Yet technical limitations to the realization of the desired potential occur from e.g. unwanted optical aberrations and speckles, which yield potential defects of small amplitude and varied length scales. 

The effect of speckle potentials on quantum systems has been of great theoretical and experimental interest due to its connection with the dirty boson concept, which deals with the consequences of the presence of impurities or disorder on a quantum assembly of interacting bosons. It has been widely studied in the case of contact interacting systems, see e.g.~\cite{Fisher1989,Huang1992,Giorgini1994, SanchezPalencia2006,Falco2007,Gaul2011,Khellil_2016,Khellil_2017, Nagler_2020} and more recently extended to the dipolar case~\cite{Pelster2011dbe,Pelster2013dbe,Pelster2014bto,Boudjemaa2015_3D,Boudjemaa2015_2D,Boudjemâa_2016_2D}. While first-order effects of potential defects have been considered in works on contact-interacting systems~\cite{Gaul2011,SanchezPalencia2006}, they have so far been neglected in studies of the dipolar case, where their cancellation over ensemble average have been assumed, i.e.\,only potential defects fluctuating from one experimental realization to the other with a vanishing ensemble average have been considered. 

In the present manuscript, we investigate the effects of static random potential perturbations on BECs with strong and tunable dipolar interactions competing with contact interactions. Such static random potentials capture the effects of optical imperfections in the realization of tailorable potentials. Our goal is to understand the impact of such imperfections in experimentally relevant situations, which we exemplify with the case of highly magnetic dysprosium atoms. We focus on perturbations of the condensate density, and due to the static nature of the potential, the dominant effects are of first order. Here we only consider the mean-field stable regime corresponding to the case of a uniform superfluid, leaving aside the vast question of the effects of potential defects on self-ordered states such as droplet crystals or supersolids. In particular, we do not include Lee-Huang-Yang corrections in the Gross-Pitaevskii equation~\cite{Chomaz2022dpa,Baillie2017,Bisset2019,Ripley2023}.

The paper is organized as follows: In Sec.\,\ref{sec:Theory} we present our model of a quasi-2D dipolar BEC and the first-order perturbation of the wavefunction for arbitrary potentials. The relevant energy scales are also discussed. In Sec.\,\ref{sec:momenta}, {we consider the simpler, yet insightful and experimentally relevant case of a static and purely cosinusoidal perturbation of the potential. This in particular describes a perturbation dominated by one Fourier component and allows to grasp the effect of one such component in a more complex perturbation.   In Sec.\,\ref{sec:white noise} we examine the distinct case of static potential perturbation where many Fourier components contribute equally, namely one characterized by a white-noise spectrum on a fixed momentum range. Such a potential defect is expected to describe the optical aberrations when forming a uniform trap. In both cases, we discuss the role of the dipolar interactions in the resulting density perturbation through its dependencies with the gas parameters.} In Sec.\,\ref{sec:conclusion}, we discuss our findings and conclude.

\section{Theoretical Model}\label{sec:Theory}

\subsection{The quasi-2D dipolar Bose gas}\label{subsec:model}

We consider a quantum Bose gas in which the particles interact via both contact interactions and DDI, and in which the dipoles are polarized by an external field. In this case, the three-dimensional (3D) interparticle interaction pseudopotential is~\cite{YiYou2001,Chomaz2022dpa}
\begin{equation}
\label{eq:Uint3D}
\Uint^{(3D)}(\bR)=\frac{4\pi \hbar^2 \as}{m}\delta(\bR)+ \frac{3\hbar^2 \add}{ m}\frac{1-3\cos^2\theta}{|\bR|^3},
\end{equation}
with $\bR$ the 3D interparticle distance, $\theta$ the angle between the dipole orientation and the interatomic axis, $m$ the particle mass, and $\as$ and $\add$ the contact and dipolar length respectively. In the case of magnetic dipoles of moment $\mu_d$,  $\add= \nicefrac{m\mu_0\mu_d^2}{12\pi\hbar^2}$ with $\mu_0$ the Bohr magneton. We define $\edd = \nicefrac{\add}{\as}$ and $g=\nicefrac{4\pi \hbar^2 \as}{m}$.

\begin{figure}
    \includegraphics[width = 0.45\textwidth]{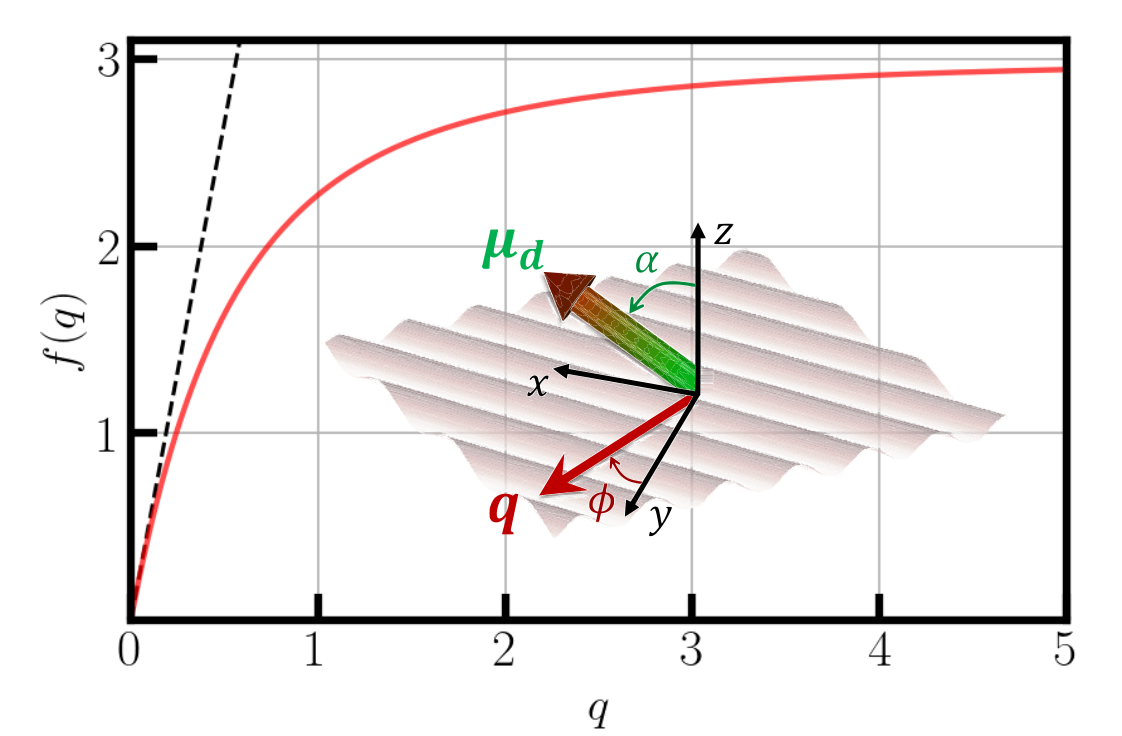}
    \caption{\label{Fig0} \textbf{Quasi-2D dipolar system and interaction} Inset: Sketch of the geometry. The gas is strongly confined along $z$, yielding a quasi-2D system in the $x,y$-plane. The dipoles (green and red arrow) are oriented in the $x,z$ plane and make an angle $\alpha$ with the $z$ axis. We typically reason in momentum space and consider a normalized wavevector $\vq$ (red arrow and red wave pattern in the $x,y$ plane) making an angle $\phi$ with the $y$ axis.  Main graph: plot of the function $f$ (Eq.\,\eqref{eq:fq}). The dotted black line is the asymptotic limit $f(q)\sim_{q\rightarrow 0} 3\sqrt{\pi}q$.}
\end{figure}

We assume that the quantum Bose gas is weakly interacting and confined in a small region of space. In particular, we use a tight and harmonic trap along $z$ of trapping frequency $\nu_z$ {such that the so-called quasi-2D regime is satisfied. In this regime,} the gas populates only the harmonic oscillator ground state along $z$, with typical length scale $\ell_z= \sqrt{h/m\nu_z}/2\pi$, and this direction can be integrated out of the equation of motion. The position in the transverse plane is represented by the vector $\br$ and the trapping potential in this plane is arbitrary and denoted $V(\br)$. The gas is described by its macroscopic in-plane ground-state wavefunction, $\psi(\br)$. In equilibrium, it satisfies the generalized stationary quasi-2D Gross Pitaevskii equation~\cite{Pitaevskii2016bec,Baillie2018dcg,Ticknor:2011asi,Fischer2006soq,Baillie2015}:
\begin{equation}
\label{eq:GP}
\mu\psi=\left[\frac{-\hbar^2}{2m}\Delta+V(\br) +\int d^3 r' \Uint({\mathbf r}-{\mathbf r}')|\psi({\mathbf r}')|^2 \right]\psi
\end{equation}
with $\mu$ being the gas chemical potential. The first term on the right-hand side of the equation is the kinetic term with $m$ the particle mass. The second term accounts for the effect of the confining potential. The third term describes the (binary) inter-atomic interactions with $\Uint(\br)$ the quasi-2D interparticle interaction potential. In the case of the 3D interaction Eq.\,\eqref{eq:Uint3D}, $\Uint(\br)$ is conveniently expressed in momentum space via~\cite{Ticknor:2011asi,Fischer2006soq,Baillie2015,Baillie2018dcg,Ripley2023}
\begin{equation}
\label{eq:Uint}
\tilde{U}(\vk)=\TF\{\Uint(\br)\}=\frac{g}{\sqrt{2\pi}\ell_z}\left[1 + \edd F(\vq,\alpha)\right]
\end{equation}
with  $\TF\{\bullet(\br)\}= \frac{1}{{(2\pi)^2}}\int \bullet(\br) e^{i \vk.\br} d\br$ the 2D Fourier transform operator, $\vk$ the conjugate coordinate to $\br$, $\vq=\vk\ell_z/\sqrt{2}$, and $\alpha$ the angle between the dipole and the $z$-axis. Assuming that the dipoles are in the ($x,z$)-plane, the function $F(\vq,\alpha)$ is
\begin{eqnarray}
\nonumber
F(\vq,\alpha) &&= (3\cos^2\alpha-1) \\
\label{eq:Fqa2}
&&+f(q)\left[\sin^2\alpha\sin^2\phi-\cos^2\alpha\right],
\end{eqnarray}
with $\phi$ the angle between $\vk$ and the $y$ axis and $q$ the norm of $\vq$. The geometry considered is illustrated in Fig.~\ref{Fig0} inset. Eq.\,\eqref{eq:Fqa2} isolates the momentum dependence in $F(\vq,\alpha)$, which will be crucial for our later study. We can also isolate its angle-$\alpha$ dependence as 
\begin{eqnarray}
\nonumber
F(\vq,\alpha) &&= \left(-1+f(q)\sin^2\phi\right) \\
\label{eq:Fqa3}
&&+\cos^2\alpha\left(3-f(q)\left[\sin^2\phi+1\right]\right).
\end{eqnarray}
The function $f(q)$ encompasses the momentum dependence and is 
\begin{eqnarray}
\label{eq:fq}
f(q)=3\sqrt{\pi}qe^{q^2}\erfc(q)
\end{eqnarray}
with $\erfc$ the complementary error function. The function $f(q)$ is plotted in Figure~\ref{Fig0}. It is positive for all $q \in \mathbb{R}^+$  and its asymptotic behaviors are $f(q)\sim_{q\rightarrow 0} 3\sqrt{\pi}q$
and $f(q)\sim_{q\rightarrow \infty} 3$.  These yield limits for the interaction potential itself as follows:
\begin{eqnarray}
\label{eq:Uint0}
\tilde{U}(\vk\rightarrow 0)&=&\frac{g\left[1 + \edd (3\cos^2\alpha-1)\right]}{\sqrt{2\pi}\ell_z},\\
\label{eq:Uintinf}
\tilde{U}(
\vk\rightarrow \infty)&=&\frac{g\left[1 + \edd (3\sin^2\alpha\sin^2\phi-1)\right]}{\sqrt{2\pi}\ell_z}
\end{eqnarray}

\subsection{Perturbation theory for the condensate density}\label{subsec:perturbth}

In the following, we consider a confinement potential $V(\br)= V_0(\br) + \delta V(\br)$ being a uniform box potential $V_0(\br)$ of size $L$ perturbed by potential defect $\delta V(\br)$, with $\int \delta V(\br) dr = 0$.  
We assume that the typical size of the trap $L$ is much larger than the healing length $\xi = \sqrt{\hbar^2/m\mu}$, defining the characteristic length scale over which the wavefunction varies under a perturbation, $L\gg \xi$. In this case, in the absence of potential perturbation $\delta V=0$, the 2D condensate wavefunction can be assumed to be uniform  $\psi(\br)=\psi_0 =\sqrt{n_0}$ with $n_0=N/L^2$ the 2D atomic density and $N$ the atom number. 
Using Eqs.\,\eqref{eq:GP},\,\eqref{eq:Uint0}, the chemical potential reads \begin{equation}
\label{eq:mu}
\mu=\tilde{U}(0)n_0=\frac{g n_0}{\sqrt{2\pi}\ell_z}\left[1 + \edd (3\cos^2\alpha-1)\right]
\end{equation}
\text

We now consider the effect of the potential defects $\delta V$ in perturbation and consider the first order expansion $\psi(\br)=\psi_0+\psi^{(1)}(\br)$, with $\int\psi^{(1)}(\br)d\br = 0$. {We then perform straightforward expansions, in a similar spirit to refs.~\cite{Pelster2011dbe,SanchezPalencia2006}.}
We prefer reasoning in momentum space and write $\psi(\vk)=\psi_0 \delta(k)+\tilde{\psi}^{(1)}(\vk)$ with $\tilde{\psi}^{(1)}(\vk)=\TF\{\psi^{(1)}(\br)\}$ 
and $\tilde{\psi}^{(1)}(0)=0$.
We also characterize the potential defect $\delta V$ via its Fourier components $\tilde{\delta V}(\vk)=\TF\{\delta V(\br)\}$. 

Expanding the Gross-Pitaevskii Eq.\,\eqref{eq:GP} to first order using  Eq.\,\eqref{eq:Uint} yields:
\begin{equation}
\label{eq:psik}
\tilde{\psi}^{(1)}(\vk)=\frac{-\sqrt{n_0}\tilde{\delta V}(\vk)}{\frac{\hbar^2\vk^2}{2m}+2n_0\tilde{U}(\vk)}.
\end{equation}
 The perturbed density at first order then writes:
\begin{eqnarray}
\label{eq:nr}
n (\br) &=& |\psi(\br)|^2 = n_0 + 2\sqrt{n_0} \int \psi^{(1)}(\vk) e^{-i \vk.\br}d\br
\end{eqnarray}
This is $n=n_0+n^{(1)}(\br)$ with $n^{(1)}$ the first order correction in the density, from which the momentum dependence is easily expressed as
\begin{eqnarray}
\nonumber
\nk(\vk) &&= \TF\{n^{(1)}(\br)\}= \frac{-2n_0\tilde{\delta V}(\vk)}{\frac{\hbar^2\vk^2}{2m}+2n_0\tilde{U}(\vk)} \\
\label{eq:nk}
&&=\frac{-2n_0\tilde{\delta V}(\vk)}{\frac{\hbar^2\vk^2}{2m}+\frac{2n_0g}{\sqrt{2\pi}\ell_z}\left[1 + \edd F(\vq,\alpha)\right]}.
\end{eqnarray}
 We introduce the dimensionless interaction parameter $p=\sqrt{32\pi}n_0a_sl_z$, which is the ratio of the contact interaction energy to the kinetic energy for $q=1$ (see Sec.~\ref{subsec:energy}). This yields
\begin{eqnarray}
\label{eq:nq}
\frac{\nk(\vq)}{n_0} = -\frac{2\tilde{\delta V}(\vq)/h\nu_z}{\vq^2+p\left[1 + \edd F(\vq,\alpha)\right]}. 
\end{eqnarray}
The equation Eqs.\,\eqref{eq:nk},\eqref{eq:nq} set the basic framework for our study. Note that this relates to the static density response function in linear-response theory, see e.g.~\cite{Pitaevskii2016bec,Bisset2019}.

In the following, we will illustrate our study using gas parameters close to those realized in experiments with highly magnetic dysprosium, namely $m=164\,$a.u. and $\add \approx \as \approx 130\,a_0$. We typically use $n_0=125\,\mu{\rm m}^{-2}$ corresponding to $5\times 10^4$ atoms trap in a box of size $L=20\mu{\rm m}$. Typical values for the axial trapping frequencies are $\nu_z=1\,$kHz$\,\approx k_B/h\times 50$\,nK corresponding to $l_z\approx 0.25\mu{\rm m}$ and   $p\approx 2$, and $\nu_z=16\,$kHz$\,\approx k_B/h\times 800$\,nK corresponding to $l_z\approx 0.06\mu{\rm m}$ and   $p\approx 0.5$. In experiments, typical in-plane trap depth are of the order of $V_0\approx k_B \times 100\,$nK to $k_B \times 1000\,$nK allowing to confine gas of temperature $T\approx 10$ to $100\,$nK. In the following, we choose as an exemplary case $V_0=h\nu_z$, which simplifies our consideration (see Eq.\,\eqref{eq:nq}) and at the same time allows to satisfy the thermal quasi-2D condition. We also choose a typical magnitude of the potential defects $\delta V$ of 1\% of $V_0$ (see also discussion in Sec.\,\ref{sec:white noise}). 
The chemical potential depends on the interaction and axial trap parameters. For $\edd=1$ and $\nu_z=1\,$kHz (16\,kHz), it varies from $\mu\approx h\times 3\,$kHz (13\,kHz) to 0 when varying from $\alpha=0$ to $\pi/2$. 
The validity of the strict quasi-2D condition thus depends on the chosen parameters.  
We note however that the quasi-2D treatment presented here extends beyond the range of validity of the strict quasi-2D condition  ($\mu,k_BT\ll h\nu_z$). Indeed the Gaussian ansatz for the transverse wavefunction has been found to apply in more generic anisotropic trap configurations, see e.g. refs.~\cite{Baillie2015,Ripley2023}.

\begin{figure}[ht!]
    \includegraphics[width = 0.5\textwidth]{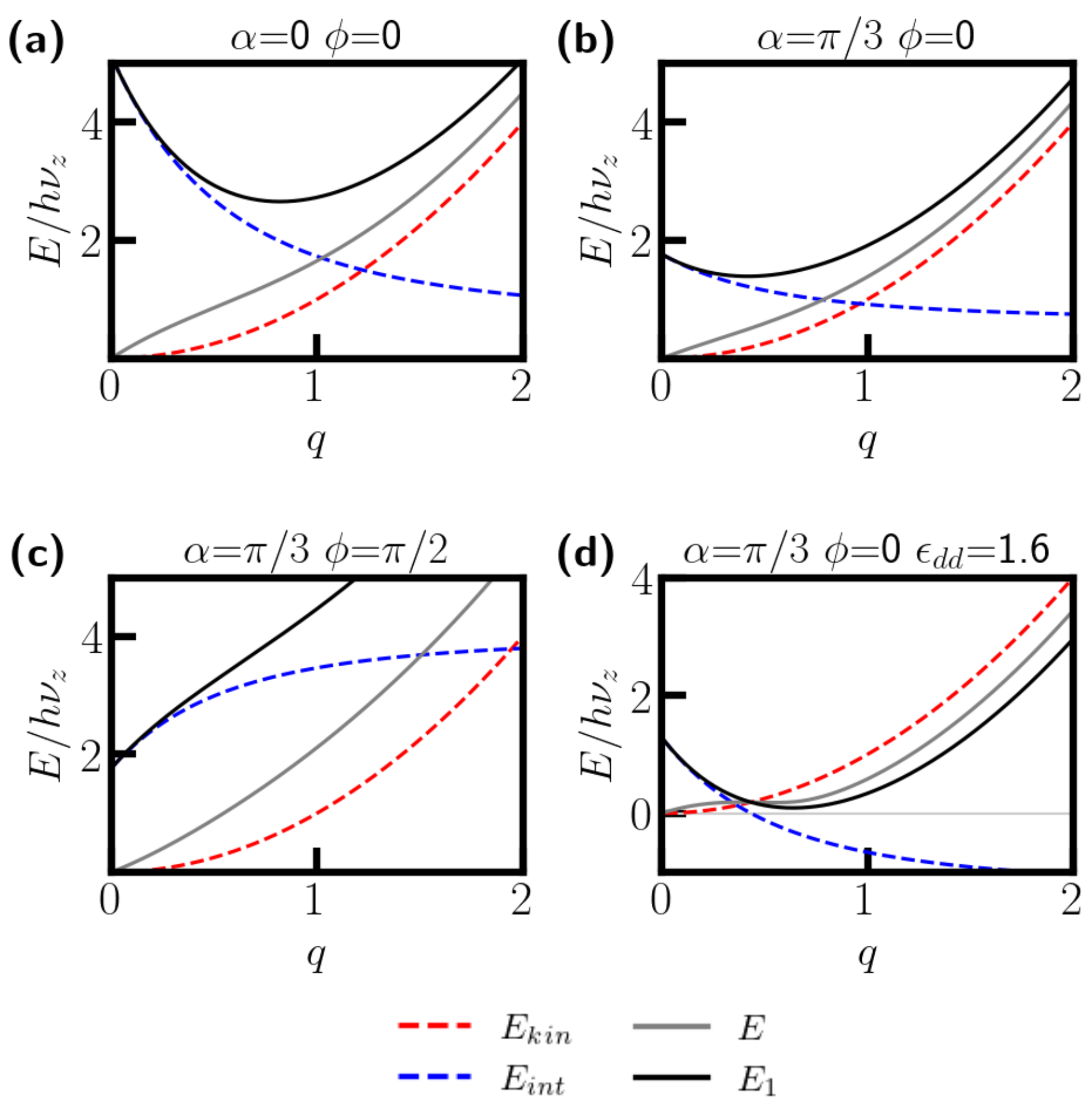}
    \caption{\label{Fig1} \textbf{Energy scales}. Values of the different energies as a function of $q$ for different angles $(\alpha,\phi)$ and values of $\edd$. The red dashed line is the kinetic term $E_{\rm kin} (\vk)$, the blue dashed line is the interaction term $E_{\rm int}(\vk)$. The black curve is the total energy $E_1(\vk)$ while the gray one is the Bogoliubov perturbation $E(\vk)$. If not specified, $\edd=0.7$. The other parameters used are $\nu_z=1\,$kHz, $n_0=125\,\mu{\rm m}^{-2}$, $m=164\,$a.u. and $a_s=130\,a_0$.}
\end{figure}

\subsection{Energy scales and their competitions}\label{subsec:energy}

Before exploring the characteristic density defects, we here comment on the characteristic energy scales appearing in our theory. In Eq.\,\eqref{eq:nk}, the characteristic energy associated to the momentum $\vk$, 
\begin{equation}
\label{eq:E1k}
E_1(\vk)=\frac{\hbar^2\vk^2}{2m}+2n_0\tilde{U}(\vk),
\end{equation}
appears in the denominator of the right-hand side. It is the sum of a kinetic $E_{\rm kin}=\hbar^2\vk^2/2m$ and an interaction term $E_{\rm int}=2n_0\tilde{U}(\vk)$. It intimately relates to Bogoliubov theory yet differs from the elementary excitation energy which is given by~\cite{Chomaz2022dpa,Pitaevskii2016bec,Baillie2015,Blakie2012}:
\begin{equation}
\label{eq:Ek}
E(\vk)=\sqrt{\frac{\hbar^2\vk^2}{2m}\left(\frac{\hbar^2\vk^2}{2m}+2n_0\tilde{U}(\vk)\right)}.
\end{equation}
In particular $E(\vk)$ has a larger contribution from the kinetic term than $E_1(\vk)$ and reads $E(\vk)=\sqrt{E_{\rm kin}(\vk)E_1(\vk)}$. This makes $E_1(\vk)$ more sensitive to interaction effects, see also~\cite{Blakie2012,Bisset2019}.

Figure~\ref{Fig1} shows the different energy scales $E_{\rm kin}$, $E_{\rm int}$, $E_1$ and $E$, as a function of $q$ in a few exemplary situations. 
At large $q$, we observe that $E_{\rm kin}$ dominates over $E_{\rm int}$ and both $E_1$ and $E$ tend to follow the variations of $E_{\rm kin}$. Instead, at small and intermediate momenta, the variations of $E_1$ and $E$  are clearly distinct. 
At small momenta, $E_1$ tends to follow the behavior of $E_{\rm int}$, taking a finite value at $\vk=0$ and exhibiting an initial variation dictated by the choice of angles $(\alpha,\phi)$.  For choices such that $E_{\rm int}$ initially decreases with $q$, $E_1$ presents a minimum at finite momentum. 

Instead, the Bogoliubov dispersion $E$ always tends to zero at $q\rightarrow 0$ and grows linearly at small momenta, following a phononic behavior. While $E_1$ is nonmonotonic in a wide range of situations, $E$ typically is monotonic. 
In some cases, $E$ becomes nonmonotonic, forming a local maximum and a local minimum at finite $q$, as observed in Fig.\,\ref{Fig1}(d). This is the celebrated maxon-roton dispersion relation~\cite{Chomaz2022dpa,Santos:2003,ODell:2003,Pitaevskii2016bec}. While the softening of a roton mode is a signature of the special dependence of the interaction energy with $q$ and of its competition with the kinetic energy, signatures of this competition are more obvious in the behavior of $E_1$ itself. Based on Eq.\,\eqref{eq:nk}, similar strong signatures on the density perturbations induced by potential defects are expected, see also~\cite{Blakie2012,Bisset2019} and later discussion.  

We note that $E_1(\vk)$ may vanish in some parameter regimes. This yields divergingly large perturbations in Eq.\,\eqref{eq:nq} and the breakdown of the underlying perturbation theory. This regime corresponds to that of so-called mean-field instability, which has been widely studied in dipolar gases in particular, see e.g.~\cite{Pitaevskii2016bec,Chomaz2022dpa,Parker:2009,Bohn:2009,Fischer2006soq} and Appendix~\ref{subsec:instab}. In this work, we restrict our study to the mean-field stable regime in which Eq.\,\eqref{eq:nq} holds. In Appendix~\ref{subsec:instab}, we quantitatively identify this regime of validity as a function of the gas parameter based on an analysis of $E_1(\vk)$.

\begin{figure*}[ht!]
    \includegraphics[width = 1\textwidth]{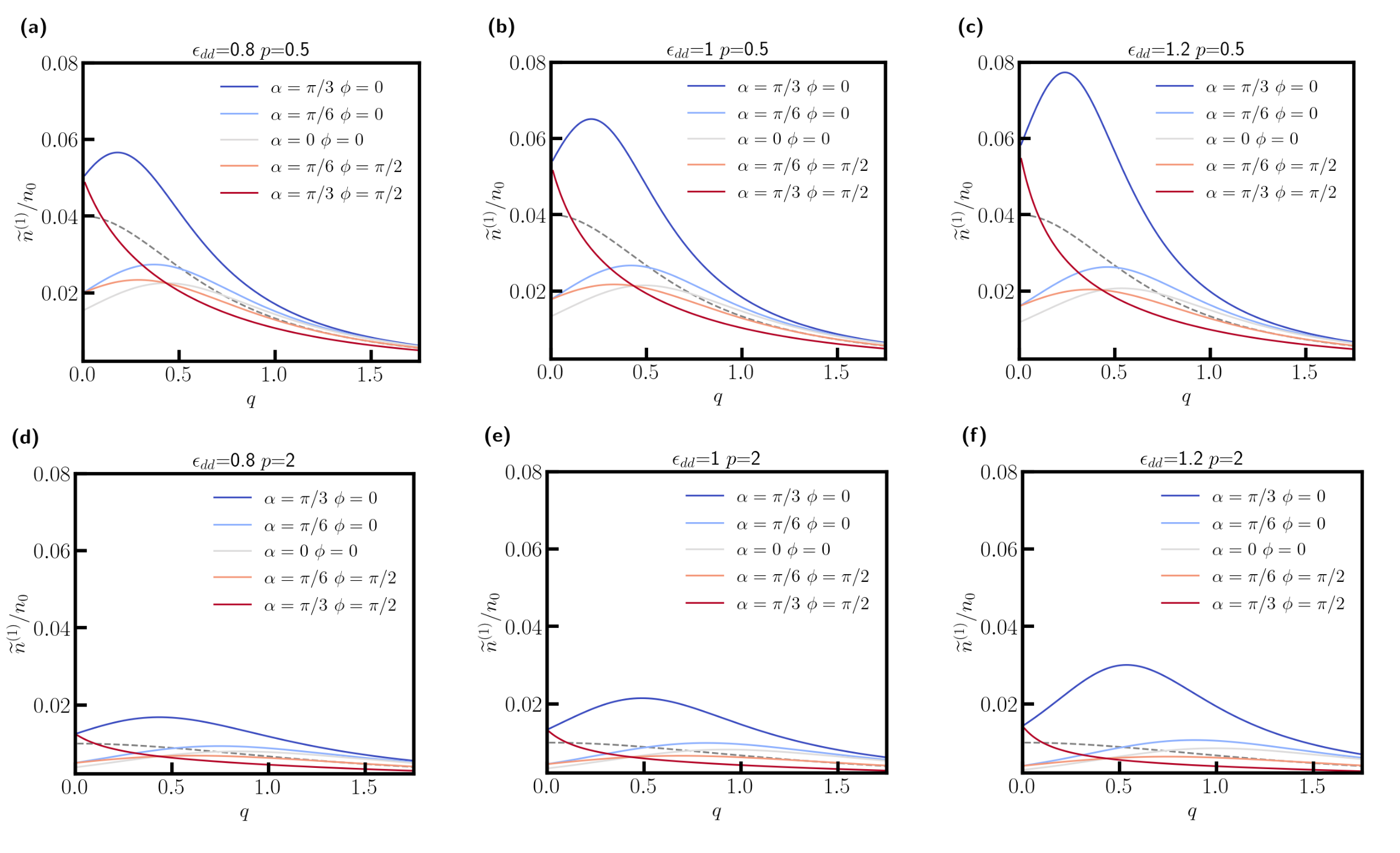}
    \caption{\label{Fig3} \textbf{Density perturbation for varying angle $\alpha$ and $\phi$}. Relative amplitude of the density perturbation $|\nk|/n_0$ for a pure cosinusoidal perturbation of the potential $\delta V=\delta V_0\cos(\vk\cdot\br)$ with an amplitude $\delta V_0$ of 1\% of the trap depth $V_0$ set to $V_0=h\nu_z$. 
    Each curve is labeled by the value of the angle $\alpha$ of the dipoles with the $z$ axis, and of the angle $\phi$, see legend. The values of the angles are chosen such that all the curves stay in the stability region, see Appendix~\ref{subsec:instab} and Fig.~\ref{Fig2}. The gray dotted curve is calculated with the same parameters but in the pure contact case, $\edd=0$. Each panel corresponds to different values of $\edd$ (0.8, 1 or 1.2) and $p$ (0.5 and 2), see panel title and discussion in Sec.~\ref{subsec:perturbth}.}
\end{figure*}
\section{Density perturbations for potential defects of fixed momenta}\label{sec:momenta}

As a first step in our analysis of potential defects, 
we consider the simple case of a pure cosinusoidal perturbation of the potential of wavevector $\vk$ with $\delta V=\delta V_0\cos(\vk\cdot\br)$. We examine the density perturbation at first order using Eq.\,\eqref{eq:nq}. This is also a cosinusoidal perturbation of wavevector $\vk$ with an amplitude given by $\nk(\vk)$, which depends on the norm of the wavevector $k$ and of the angle of excitation direction $\phi$. {The study of the cosinusoidal case is both instructive and experimentally relevant: It allows an easy understanding of the role of the different system parameters on the strength of the density perturbation. Experimentally cosinuoidal perturbations are easily realized by creating, either intentionally or unintentionally, a standing wave at the position of the atoms, forming a so-called optical lattice, see e.g.~\cite{Pitaevskii2016bec,Chomaz2022dpa}. As we will see in more detail in Sec.~\ref{sec:white noise}, a cosinusoidal perturbation, through Fourier decomposition, also describes the effect of one part of a more complex perturbation. While in Sec.~\ref{sec:white noise}, we will investigate the case of a perturbation where many momentum contributions have the same weight, it can instead occur that the experimental potential perturbation has contributions of unequal weights and is dominated by a given momentum. Understanding the interplay of a given contribution with the different parameters will then allow to tailor the response of the system, minimizing or enhancing the density perturbation generated by that contribution. Below, we examine the norm $|\nk(\vk)|$ of the cosinusoidal density perturbation.} 

\begin{figure*}[ht!]
    \includegraphics[width = 1\textwidth]{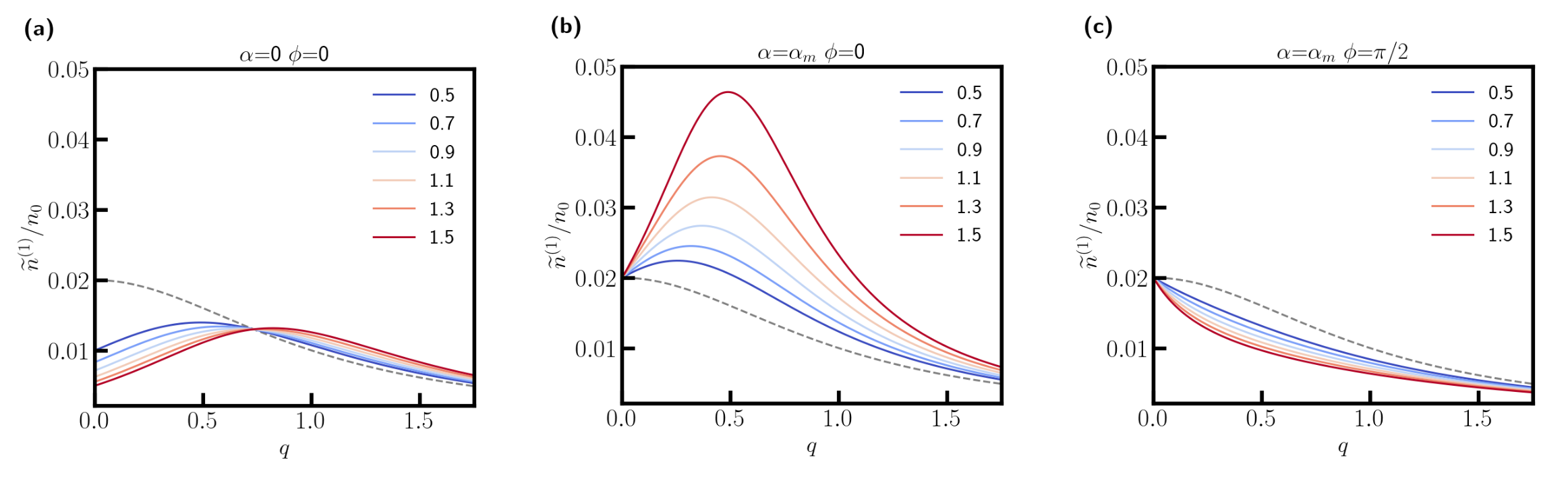}
    \caption{\label{Fig4} \textbf{Density perturbation for varying $\edd$}. Relative amplitude of the density perturbation, $|\nk|/n_0$ for a pure cosinusoidal perturbation of the potential $\delta V=\delta V_0\cos(\vk\cdot\br)$ with an amplitude $\delta V_0=1\% V_0$ with $V_0$ set to $V_0=h\nu_z$. 
    Each curve is labeled by the value of $\edd$ (legend). The gray dotted curve is calculated with the same parameters but in the pure contact case, $\edd=0$. Each panel is for a different set of angles, see panel title.}
\end{figure*}

In the nondipolar case $\tilde{U}(\vk)=g/\sqrt{2\pi}\ell_z$ such that $|\nk(\vk)|$ is independent on $\phi$ and is a Lorentzian in $k$ with a maximum at $k=0$ and a characteristic momentum width given by that associated with the healing length. Adding dipolar interactions modifies $|\nk(\vk)|$ in a nontrivial way (see Eq.\,\eqref{eq:Uint}). 
Yet, using Eqs.\,\eqref{eq:Uint0}-\eqref{eq:Uintinf}, we find that the perturbation effects at $k=0$ and $k \rightarrow \infty$ behave similarly to the pure contact case. At $k=0$, $|\nk|$ has a finite amplitude independent of $\phi$. In the large $k$ limit,  $|\nk|$ vanishes quadratically in $k$ as the kinetic term dominates in the denominator of Eq.\,\eqref{eq:nq}. The dipolar effects thus reveal at small but finite values of $k$.
We here explore these effects through a numerical study of the density perturbation as a function of $k$, for different $\vk$ orientation (angle $\phi$, see Fig.~\ref{Fig0} inset), and different gas parameters, namely $\alpha$, $\edd$, $p$, see Figs.\,\ref{Fig3} and \ref{Fig4}.  

A first stringent effect of DDIs is evident in Fig.\,\ref{Fig3} through the fact that, for most sets of parameters, $|\nk|$ is nonmonotonic and has a maximum at a nonzero momentum, $\km$. This differs from the pure contact case where the maximum is always at $k=0$.  Even in cases where a maximum is found at $k=0$ (see $\alpha=\pi/3$ and $\phi=\pi/2$), the behavior of $|\nk(k)|$ differs from the contact case, in particular it exhibits a nonvanishing derivative at $k=0$. These behaviors directly relates to those of $E_1(k)$ (Eq.~\eqref{eq:E1k}) discussed in Sec.~\ref{subsec:energy} and Fig.~\ref{Fig1}. In particular, the nonmonotonic $|\nk(k)|$ results from the emergence of a minimum in $E_1(k)$, i.e. from an interaction contribution that decreases with $k$ (dominance of long-range attraction at small wavelengths) and that dominates over the kinetic contribution up to $k\approx \km$. 
The nonmonotonic behavior of $|\nk|$ indicates enhanced density fluctuations for perturbations at $k\approx\km$. Characterizing $\km$ and its variations with the perturbation direction and gas parameters is thus crucial to apprehend the density perturbations arising from potential defects in dipolar gases. Its variations with  $\alpha$, $\edd$ and $p$ can be observed in Fig.\,\ref{Fig3}.
In the individual panels, we observe that, for all $\phi$ values, $\km$ shifts to lower values when increasing $\alpha$. This is because increasing $\alpha$ reduces the interaction energy at small momenta (by increasing the dipolar attraction contribution in plane) and thus shifts the dominance of the kinetic energy term in $E_1$ to lower $k$. Comparing panels of respectively one row and one column in Fig.\,\ref{Fig3}, we also observe that, for all $\alpha$ and $\phi$, $k_m$ increases with both $p$ and $\edd$. This is because increasing $\edd$ and $p$ increases the overall strength of the dipolar interaction term and shifts its competition with the kinetic term in $E_1$ to large momenta. {This analysis is of experimental relevance as, in the presence of a potential perturbation of given wavelength, tuning the interaction parameters and angles, as well as the transverse trapping scale (which rescales the momenta), may allow to enhance or mitigate the impact on the density of the system through an appropriate shift of $\km$.}

Another effect of the DDI anisotropy is evidenced by comparing the curves at $\phi=0$ (perturbation occurring perpendicular to the dipole plane) and $\pi/2$ (perturbation in the dipole plane) for a fixed set of gas parameters in each panel of Fig.~\ref{Fig3}. Differences in $|\nk|$ occur between these two perturbation directions. They diminish with decreasing $\alpha$, and disappear at $\alpha=0$, matching the evolution of the DDI in-plane anistoropy. First, $|\nk|$ is always larger for $\phi=0$ than for $\phi=\pi/2$. This effect connects to the additional repulsive contribution $\propto \sin^2\alpha \sin^2\phi $ in the momentum-dependent interaction of Eq.~\eqref{eq:Fqa2} (second line) for $\phi>0$, coming from interaction perpendicular to $\vk$, see discussion below. {Experimentally, this allows to reduce (or enhance) the density perturbations by orienting the dipoles along (or perpendicular to) the momentum of the dominant contribution in $\delta V$.}

Second, while in the case $\phi=0$, $|\nk|$ increases with $\alpha$ for all $k$, in the case of $\phi=\pi/2$, its behavior with $\alpha$ depends on $k$: $|\nk|$ increases (resp.\,decreases)  with increasing $\alpha$ for small (resp.\,large) $k$. Remarkably all curves with $\phi=\pi/2$ intercepts at the same $k=k_*$ in each panel of Fig.~\ref{Fig3} meaning that the value of the momentum at which the change of behavior occurs, $k_*$, is independent on $\alpha$ for a fixed $p$ and $\edd$. This can be understood from Eq.~\eqref{eq:Fqa3} where the $\alpha$-dependent contribution is isolated in the second line as $\cos^2\alpha (3-f(q)(\sin^2\phi+1))$. This contribution is repulsive at small $k$, attractive at large $k$ and changes sign for $q_*=k_*l_z/\sqrt{2}$ solution of $f(q_*)=\frac{3}{(\sin^2\phi+1)}$ independent on $\alpha$. For $\phi=\pi/2$, $q_*\approx 0.43$ and thus corresponds to $k_*\approx 1/l_z$. For $\phi=0$, the equality reads $f(q_*)=3$, yielding $q_*\rightarrow\infty$. The various behaviors with $\alpha$ can thereby be understood as a competition of the length scales setting the interaction cost of a density modulation of wavevector $\vk$. Due to the long-range nature of the interaction, this cost has contributions from interactions not only in $\vk$ direction, but also in the direction transverse to $\vk$, the weight of the latter increasing with $k$. For $\phi=0$, increasing $\alpha$ always increases the role of dipolar repulsion. Instead for $\phi=\pi/2$, the behavior depends on $k$, and for $k>k_*\sim 1/l_z$, the interactions along $z$ become dominant, thus inverting the dependence with $\alpha$ compared with the case $k<k_*$. {Experimentally, this observation allows to tune the strength of the density perturbation with $\alpha$ by changing the orientation of the dipole compared with the momentum of the potential perturbation, and by tuning the strength of the transverse confinement, which sets the scale for $k_*$.}

Besides the angular dependence, the development of the density perturbation is also influenced by the gas parameters, which are tunable in experiments, in particular the relative strength of contact and dipolar interaction $\edd$, the overall strength of the interactions set by $\as$, the mean density $n_0$, and the axial confinement strength $\nu_z$. The first parameter enhances the dipolar effects and its tuning is typically exploited in experiments~\cite{Chomaz2022dpa}. The latter two parameters rescale the density perturbation ($\nk/n_0$), the potential perturbation ($\tilde{\delta V}/h\nu_z$), and the momenta ($q =k \ell_z/\sqrt{2}$) in Eq.\,\eqref{eq:nq}. Together with $\as$, these two parameters also set the competition between the kinetic and interaction energy in Eq.\,\eqref{eq:E1k} as encompassed by the dimensionless parameter $p \propto n_0\as l_z$.   

The effect of varying $p$ is observed in Fig.~\ref{Fig3} by comparing panels of one column and that of changing $\edd$ by comparing the panels of one row.  We note that the effect of tuning $\nu_z$ or $n_0$ in experiments is not directly included in the $p$-depedence but also involves the rescaling of the momenta and perturbation strengths. In particular, we note that the in-trap potential defects are simply rescaled by $h\nu_z$ in Fig.~\ref{Fig3}, while they experimentally scale with $V_0$, independent of $\nu_z$. Furthermore, experimentally tuning $\as$ typically rescales not only $p$ but also $\edd$ since $\add$ is a constant, at least in the case of magnetic atoms. Yet even if $p$ is not easily independently tunable in experiments, it is an interesting quantity as the density perturbation shows a simple dependence, with $|\nk|$ decreasing for increasing $p$ for all the angular configurations. This decrease competes with the rescaling of the density and potential perturbation by $n_0$ and $h\nu_z$, respectively.

The effect of $\edd$, visible in Fig.~\ref{Fig3} by comparing the panels of one row, is instead more intricate and depends on $\alpha$ and $\phi$.  We further investigate this dependence in Fig.~\ref{Fig4} where $|\nk(k)|$ is shown for varying $\edd$ and showcased for different values of these angles. 
In the case of small $\alpha$ (see Fig.\,\ref{Fig4} (a)), the dependence on $\edd$ depends on $k$, with $|\nk|$ decreasing (resp.\,increasing) for increasing $\edd$ at small (resp.\,large) momenta. This is explained as for dipoles nearly perpendicular to the atomic plane, the contribution of the DDI is repulsive (resp.\,attractive) as small (resp.\,large) momenta, see discussion above.  The momentum $k_*'$ at which the behavior change corresponds to the value of $q$ where $F(\vq,\alpha)$ changes of sign, which is independent on $\edd$, see Eq.\,\eqref{eq:Fqa2}. For $\alpha=0$, $q_*'\approx 0.73$ yielding $k_*'\approx 1/l_z$, meaning that the transverse length set the scale below which the perturbations are amplified by increasing $\edd$.
The value of $k_*'$ varies with $\alpha$, decreasing for larger angles. This is explained as increasing $\alpha$ shifts the attractive contribution of the DDI to small momenta. We note that while at $\alpha =0$, the behavior is independent of the excitation direction $\phi$, at finite $\alpha$ anisotropic behaviors occur and $k_*'$ depends on $\phi$, see Eq.\,\eqref{eq:Fqa2}. {This analysis is of experimental relevance as tuning the transverse confinement and the dipole orientation allows to minimize or enhance the perturbations when tuning $\edd$.}

The panels \ref{Fig4}(b,c) illustrate a larger tilt of the dipoles, at the so-called magic angle $\alpha_m=\arccos(1/\sqrt{3})$, and for two different directions of excitations $\phi=0$ (b) and $\phi=\pi/2$ (c). At this angle, $k_*'$ is shifted to $k=0$ for all excitation directions, yet $|\nk|$ shows distinct behaviors for the different excitation directions at finite $k$: $|\nk|$ has distinct $k$-dependences, it is significantly larger for $\phi=0$ than for $\phi=\pi/2$, and it increases for increasing $\edd$ for $\phi=0$, while the behavior with $\edd$ is inverted for $\phi=\pi/2$.  These different variations with $\edd$ relate to the finite-$k$ dipolar contribution being attractive (resp. repulsive) for $\phi=0$ (resp.\,$\phi=\pi/2$), see second line of Eq.~\eqref{eq:Fqa2}. In this configuration, {the direction of the dipoles compared with the perturbation mostly sets the dependence on $\edd$ and can be tuned experimentally for this purpose}. We note that for the case $\alpha=\alpha_m$ and $\phi=\pi/4$, $F(\vq,\alpha)$ vanishes for all $\vq$ and hence there is no dipolar effect in the density perturbation. 

The above analysis has shown that perturbations in the potential lead to density fluctuations of variable strength depending on the parameters of the perturbation and of the gas. This can be detrimental, since deviations from the uniform case occur differently in different experimental conditions, but it can also be used to tailor the response of the system.  
{As discussed in several instances above, such dependencies can be used to e.g. minimize the effect of potential defects with a dominant momentum contribution, or instead to use the enhanced response of the system to induce pattern formations by matching the potential modulation momentum to $\km$.}

\section{Density perturbations for static white-noise potential defects}\label{sec:white noise}

{After having understood the effects of individual Fourier components of the potential defects on the density perturbations and the means of controlling them in Sec.\,\ref{sec:momenta}, we are interested in studying density perturbations induced by potential perturbations consisting of many equally contributing components. 
In this way, we aim to account for} potential defects arising from optical imperfections in the experimental realization of uniform box potentials. Uniform potentials are typically created using spatial light modulators to shape flat-top or flat-bottom beams~\cite{Atomtronic2021,Navon2021,Gauthier2021dhr}. Experimental characterizations of beams created in this kind of setup show noise with almost constant amplitude of the Fourier components, see e.g.~\cite{Wang:08}.  In our experiment, we have observed similar Fourier composition and typical standard deviation of the flat top beam on the order of $1\%$~\cite{ThibaultThesis}. In the following, we will consider random but static potential defects (meaning that there are no significant shot-to-shot fluctuations for a given experimental realization) and characterize them by the complex amplitude of their different momentum components, $\delta \tilde V (\vk)$.  
 

More precisely, we consider a potential defect $\delta V$ formed by a static white noise over momenta of norm in the range $[\kmin, \kmax]$ and all possible orientations $\phi$. We ensure {that $\delta V(\br)$ is real-valued and that} 
$\int \delta V(\br)dr=0$, and characterize $\delta V$ by its spatial standard deviation $\sigma_V=\left(\int |\delta V (\br)|^2 dr\right)^{1/2}/L$. We use a perturbation of $\sigma_V= 0.01V_0$ and set the box trap of depth to $V_0= h \nu_z$ (see Sec.~\ref{subsec:perturbth}).
We choose $\kmin=2\pi/5L$ with $L=20\,\um$, limiting the wavelength of the potential defect to a few (5) times the box size, and $\kmax=2\pi/\lambda$ with $\lambda = 500\,$nm, limiting the wavelength of the potential defect to the wavelength of the light used to generate the trap (diffraction limit). This yields $[\kmin,\kmax]=[0.06 ,12.6]\,\mu\textrm{m}^{-1}$ and $[\qmin,\qmax]=[0.011, 2.2]$ (resp. $[0.0028 ,0.55]$) for $\nu_z=1\,$kHz (resp. 16\,kHz). We define the momentum $\vk$ through its axial coordinates $(k_x,k_y)$ computed on a numerical grid of discretization $\delta k = 2\pi\times10^{-2}\,\mu\textrm{m}^{-1}$ {(we have checked that the simulation results do not depend on $\delta k$ if smaller values are used)}. A single realization of the random potential is obtained by randomly drawing the phase of each momentum component of the potential defect according to a uniform distribution, and then imposing  the potential defect's standard deviation to match $\sigma_V$. {In the random draw defining $\delta \tilde V(\vk)$, we consider $\vk$ only if $k=\sqrt{k_x^2+k_y^2}$ satisfies $k\in[\kmin,\kmax]$ and we restrict to $k_x \geq 0$, and further set $\delta \tilde V (-k_x,-k_y)=\delta \tilde V^* (k_x,k_y)$ where $.^*$ is the complex conjugate, which ensures that $\delta V(\br)$ is real. Since $\kmin>0$, $\delta \tilde V (\vk=0)=0$ and $\int \delta V(\br)dr=0$.}

\begin{figure*}
    \includegraphics[width = 1\textwidth]{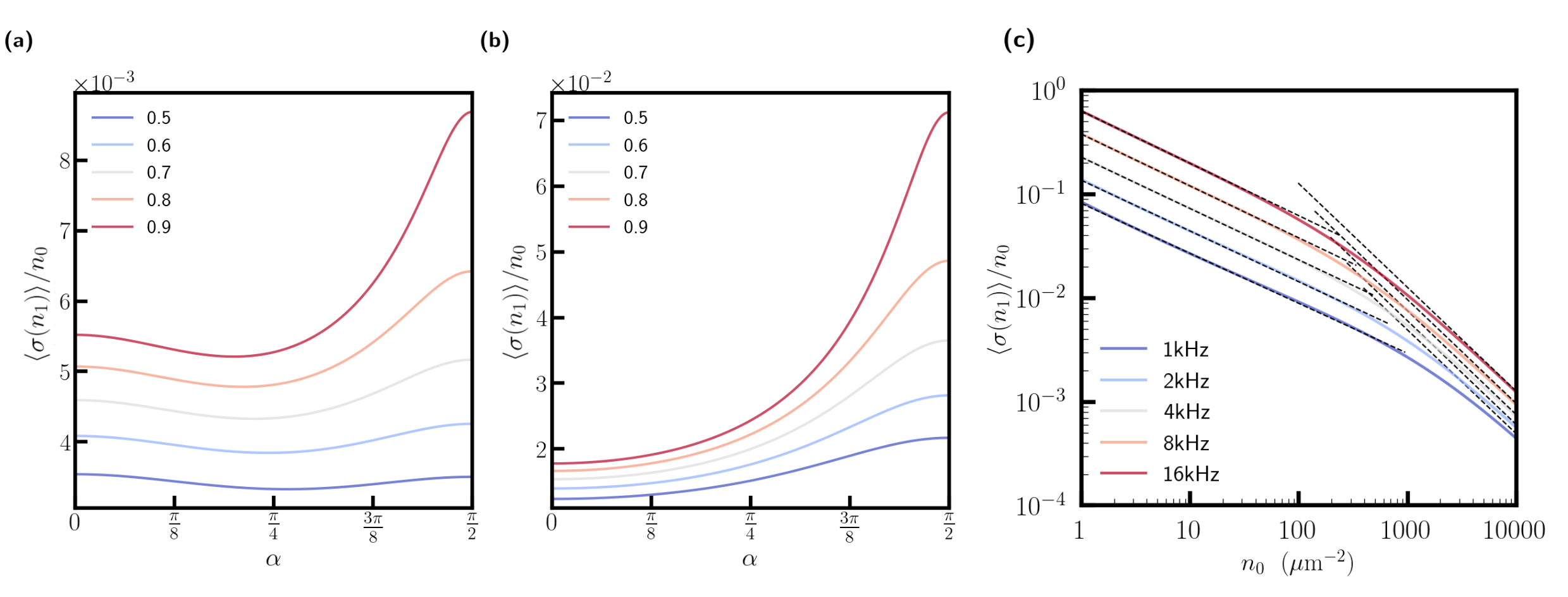}
    \caption{\label{Fig7} \textbf{Density perturbation in white-noise potential}. Relative standard deviation of the density of an atomic cloud of density for a white-noise perturbed potential with $\sigma_V= 1\% V_0$ and $V_0= h \nu_z$. 
    We set $m=164\,$a.u.,  $\add=133\,a_0$. (a,b) $\langle\sigma(n_1)\rangle/n_0$ as a function of  $\alpha$, with $\nu_z=1\,$kHz in (a) and  $\nu_z=16\,$kHz in (b). The different curves are labeled by the value of $\edd$ (legend). (c) $\langle\sigma(n_1)\rangle/n_0$ plotted in a log-log scale as a function of the background density $n_0$. The different curves are labeled by the value of $\nu_z$ (legend). The black dotted lines are linear fit of the values at high density $n_0>10\noc(\kmax)$ and low densities $n_0<0.1\noc(\kmax)$ (see text). The other parameters used are $\as=150a_0$ and $\alpha=0$.}
\end{figure*}

Based on the potential defect $\delta \tilde V (\vk)$, we obtain the spatial density perturbation at first order, $n_1(\br)$, by applying Eq.\,\eqref{eq:nk} to each momentum component and calculating the inverse Fourier transform of $\nk(\vk)$. As we are interested in static potential defects, we do not perform an ensemble average on the density perturbation itself, contrarily to previous dirty bosons studies~\cite{Pelster2011dbe,Pelster2013dbe,Pelster2014bto,Boudjemaa2015_3D,Boudjemaa2015_2D,Boudjemâa_2016_2D}. Instead, we examine the spatial standard deviation of the resulting density perturbation over a single realization of the potential defect, calculating $\sigma(n_1)=\left(\int n_1^2 (\br) dr\right)^{1/2}/L$. We then perform an ensemble average of the deviation $\langle\sigma(n_1)\rangle$  over 20 realizations of the random potentials of the same characteristics $\sigma_V$, $\kmin$, $\kmax$, see also~\cite{Gaul2011,SanchezPalencia2006}. We note that Eq.\,\eqref{eq:nk}  and $\int \delta V(\br)dr=0$ yield  $\int n_1 (\br) dr=0$.  We study the variations of $\langle\sigma(n_1)\rangle$ as a function of the different parameters $(\alpha$,\,$\edd$,\,$n_0$,\,$\nu_z)$ in Fig. \ref{Fig7}.

Figure \ref{Fig7} (a) {and (b)} shows $\langle\sigma(n_1)\rangle/n_0$  as a function of the dipole angle $\alpha$ for various values of $\edd$, and for {$\nu_z=1\,$kHz and $16\,$kHz, respectively}. We keep $\edd<1$ to avoid instabilities (see Appendix~\ref{subsec:instab}). The dependencies of the density perturbation on $\alpha$ and $\edd$ appear  relatively simple {here compared with the subtle dependencies of $\nk/n_0$ discussed in Sec.\,\ref{sec:momenta}: For a given value of $\nu_z$, a similar evolution of $\langle\sigma(n_1)\rangle$ with $\alpha$ is observed for all values of $\edd$. In the case $\nu_z =1\,$kHz (a), for which $[\qmin,\qmax]$ contains both values smaller and larger than 1, the evolution is nonmonotonic, first decreasing and then increasing. Given the observations of Sec.\,\ref{sec:momenta},  these trends imply that contributions at large $\phi$ and large $q$ values (see discussion on $q_*$) dominate the response at small $\alpha$, while the response at large $\alpha$ is dominated by other contributions, of either small-$q$ or small-$\phi$ components. The position of the minimum in $\langle\sigma(n_1)\rangle$ shifts to smaller $\alpha$ with increasing $\edd$, indicating a reduced dominance of the large $q$ and $\phi$ components at larger $\edd$. In the case $\nu_z =16\,$kHz (b), for which $[\qmin,\qmax]$ contains only values up to 0.55, the evolution is instead monotonically increasing, reflecting the fact that the potential contains only smaller $q$ contributions.}

{Comparing (a) and (b), we see that the amplitude of $\langle\sigma(n_1)\rangle$ and the strength of its variations increases with increasing $\nu_z$. This relates to the associated decrease of $p$, see Sec.\,\ref{sec:momenta} and Fig.\,\ref{Fig3}. We note that, as in the previous section, we impose $V_0=h\nu_z$, which does not necessarily holds in experiments and affects the dependence of $\langle\sigma(n_1)\rangle$ on $\nu_z$. Making $V_0$ independent on $\nu_z$ yields the opposite dependence. Furthermore, in both (a) and (b),} we observe a strong impact of the value of $\edd$ on the amplitude of $\langle\sigma(n_1)\rangle$ with overall larger amplitudes and stronger variations with $\alpha$ as $\edd$ increases. {Given the observations of Sec.\,\ref{sec:momenta} and Fig.\ref{Fig4}, this indicates a dominance of the contributions with large $q$ (see $q_*'$) at small $\alpha$ and with small $\phi$ at large $\alpha$.} In both cases, the sharper increase of $\langle\sigma(n_1)\rangle$ is found for $\alpha\simeq \pi/2$ when $\edd\rightarrow 1$, which relates to the mean-field instability discussed in Appendix~\ref{subsec:instab}, see Fig.\,\ref{Fig2}. Potential defects might then seed characteristic and sizeable density modulations in the system. We note that, once such a seeding reveals macroscopically a treatment beyond our perturbative approach is required. We also note again that in the regime of mean-field instability, the standard mean-field theory loses its validity, and beyond-mean-field effects must be included, see e.g.~\cite{Baillie2017,Bisset2019,Ripley2023}. {For $\edd =0.9$, $\langle\sigma(n_1)\rangle/n_0$ varies of $60\%$ (resp. $170\%$) of its mean value in Fig. \ref{Fig7} (a) (resp.(b)). The variations reduces for more moderate values of $\edd$, and the effect of the potential defects on the density can, in experiments, be  considered to be roughly independent of the dipoles' orientation at low enough $\edd$. A weaker transverse confinement allows to reach this condition for larger $\edd$.} 

Figure \ref{Fig7} (c) further reports on the effect of the gas parameters on the amplitude of $\langle\sigma(n_1)\rangle/n_0$, showing its variation as a function of the mean density $n_0$ and for different confinement strength $\nu_z$, for $\alpha=0$. {The dependence on $\nu_z$ shows that the observation made for Fig. \ref{Fig7} (a), (b) is valid for the whole range of densities.} 
The dependence of $\langle\sigma(n_1)\rangle/n_0$ on $n_0$, here plotted in log-log scale, evidences two distinct scaling regimes, holding respectively at high and low densities. In both cases, we observe a power scaling that appears to hold with similar power for all values of $\nu_z$. Fits of the calculated $\langle\sigma(n_1)\rangle/n_0$ to $ n_0^{-\gamma}$ for different values of $\nu_z$ yield $\gamma=0.46\pm0.02$ and $\gamma=1\pm10^{-5}$ at small and large densities respectively. {We note that a similar behavior is found for different orientation of the dipoles with the only noticeable difference in the power law scaling holding at low density, which is fitted to $\gamma=0.59\pm0.02$ for $\alpha=\pi/2$.}   

These different behaviors as a function of $n_0$ relate to energy competition in $E_1(k)$ (Eq.\,\eqref{eq:E1k}) between the density-independent kinetic and the density-dependent interaction terms. Kinetic and interaction terms dominate at large and small momenta respectively.  
For a perturbation of a given momentum $k$, the density $\noc(k)$ at which the interaction and kinetic terms are of the same order of magnitude is approximately given $\noc(k)\sim\frac{k^2l_z}{8\sqrt{2\pi}\as}$. We can thereby introduce two characteristic densities: $\noc(\kmax)$ above which the interaction term dominates for all the momenta of our perturbed potential and $\noc(\kmin)$ below which the kinetic term dominates even for the smallest momentum of our perturbation. For $\noc(\kmin)<n_0<\noc(\kmax)$, an intermediate behavior occurs where the kinetic term dominates for the large momenta while the interaction dominates at small momenta. In our setting and using $\nu_z=1\,$kHz, we find $\noc(\kmin)\approx 6\times 10^{-3}\,\um^{-2}$ and $\noc(\kmax)\approx 2\times 10^2\,\um^{-2}$. 
For $n_0>\noc(\kmax)$, the dominance of interaction effects yields $\frac{\nk(\vk)}{n_0}\simeq \frac{-\tilde{\delta V}(\vk)}{n_0\tilde{U}(\vk)}$ which implies a perturbation $n_1$ independent of the background density $n_0$. This behavior matches the fitted scaling at large densities in Fig.\,\ref{Fig7} (b). For $n_0<\noc(\kmin)$, the dominance of kinetic effects yields a normalized perturbation $\frac{n_1}{n_0}$ independent of density. Such a low-density regime is not observed in Fig.\,\ref{Fig7} (b). {In the intermediate regime $\noc(\kmin)<n_0<\noc(\kmax)$, the intermediate scaling ${\nk(\vk)} \sim n_0^{\gamma}$, $0<\gamma \approx 0.5 <1$ observed may be explained by partitioning the momentum range $[\kmin, \kmax]$ in kinetic and interaction dominated range with a density-dependent momentum cut $k(n_0) =\kappa \sqrt{n_0}$ and $\kappa=(2\pi)^{1/4}\sqrt{\frac{8a_s}{l_z}}$. This partition shall lead an intermediate scaling behavior.} 
We stress that the observed density dependence is dictated by the range of momenta contained in the perturbation potential and in particular by the value of $\kmax$, which set $\noc(\kmax)$. From an experimental point of view, it is then interesting to correct the small-scale defects of the potential in order to decrease $\kmax$, and thus $\noc(\kmax)$. The high-density scaling regime discussed above is reached at lower densities, which is advantageous since this regime exhibits stronger suppression of the density perturbation.

\section{Conclusions and perspectives}\label{sec:conclusion}

In this work, we have investigated the density perturbations of a quasi-2D dipolar Bose gas in box potentials perturbed by static aberrations. 
We first studied {perturbing potentials formed or dominated by a single momentum component}. We explored the dependence of the density perturbation on this momentum and on the gas and interaction parameters. The anisotropy of the dipolar interactions gives rise to angular dependencies, and its long-range character translates into effects at intermediate momenta. For a wide range of parameters, the perturbation exhibits a maximum at a finite wavenumber whose value varies with the gas parameters, indicative of enhanced fluctuations at this finite momentum. The observed variations of the density perturbations with the gas and interaction parameters need to be taken into account in experiments. They may also allow to tailor the density perturbations in a perturbed trap by tuning the parameters to minimize or maximize the response of the system at the perturbing momentum. Besides the interaction parameters, the transverse trapping frequency and the mean 2D density are also useful parameters to tune.

We then studied the density perturbation induced by a {perturbing potential characterized by a static white-noise spectrum} on a given range of wavenumbers. Here again, the anisotropy of the dipolar interactions induces angular dependencies whose strength is found to increase with increasing dipolar strength $\edd$ and transverse trapping frequency $\nu_z$. Additionally, variations of the transverse trapping frequency and the mean 2D density are found to enhance and suppress the relative fluctuations of the density, respectively. The density-induced suppression operates with characteristic scaling laws dependent on the mean density range. At high densities, a favorable interaction-dominated regime is found with $n_1/n_0 \propto 1/n_0$ while at intermediate lower densities we observe $n_1/n_0 \propto 1/n_0^\gamma$ {where the exponent $\gamma\approx 0.5$ depends on $\alpha$}. The transition density between these two regimes is fixed by the momentum composition of the potential defect, and in particular dictated by the lowest momenta contained in the perturbation. Correcting for defects at small momenta is therefore favorable to minimize the density defects. 

Our work explores the possibility of realizing dipolar quantum gases confined in uniform potentials. It sheds light on the effect of unavoidable potential imperfections, e.g. due to optical aberrations on the dipole trapping beam. It highlights the challenge of realizing homogeneous dipolar gases over a wide range of gas parameters, but also the possibility of mitigating the potential imperfections by a careful choice of parameters and dedicated corrections. 

Here we have considered only first-order perturbations in the mean-field stable regime. Our work can be extended to account for second-order effects as in \cite{Pelster2011dbe,Pelster2013dbe,Pelster2014bto,Boudjemaa2015_3D,Boudjemaa2015_2D,Boudjemâa_2016_2D} where a condensate depletion is predicted. Furthermore, another extension of the present work will be to address the regime of mean-field instability, which could be done, for example, by including a beyond-mean-field stabilization term in the GPE~\cite{Baillie2017,Bisset2019,Ripley2023,Chomaz2022dpa}. This extension would allow to explore the intriguing effects of (involuntary or voluntary) potential defects on self-organized ground states (e.g. supersolids).

\begin{acknowledgments}

We thank Karthik Chandrashekara, Jianshun Gao, Shuwei Jin, Christian Gölzhauser, Charles Drevon, Sarah Philips, Maurice Rieger, and Britta Bader for insightful discussions. We thank Wyatt Kirkby for his careful reading and helpful comments on the manuscript. This work is funded by the European Research Council (ERC) under the European Union’s Horizon Europe research and innovation program under grant number 101040688 (project 2DDip), and by the Deutsche Forschungsgemeinschaft (DFG, German Research Foundation) through project-ID
273811115 (SFB1225 ISOQUANT) and under Germany’s Excellence Strategy EXC2181/1-390900948 (the Heidelberg Excellence Cluster STRUCTURES). Views and opinions expressed are however those of the authors only and do not necessarily reflect those of the European Union or the European Research Council. Neither the European Union nor the granting authority can be held responsible for them.  
\end{acknowledgments}
\vspace{5pt}

$\star$ Correspondence and requests for materials should be addressed to chomaz@uni-heidelberg.de.

\appendix
\renewcommand\thefigure{\thesection S\arabic{figure}}   
\setcounter{figure}{0}   

\section{Mean-field stability analysis based on perturbation theory}\label{subsec:instab}

\begin{figure}[ht!]
    \includegraphics[width = 0.5\textwidth]{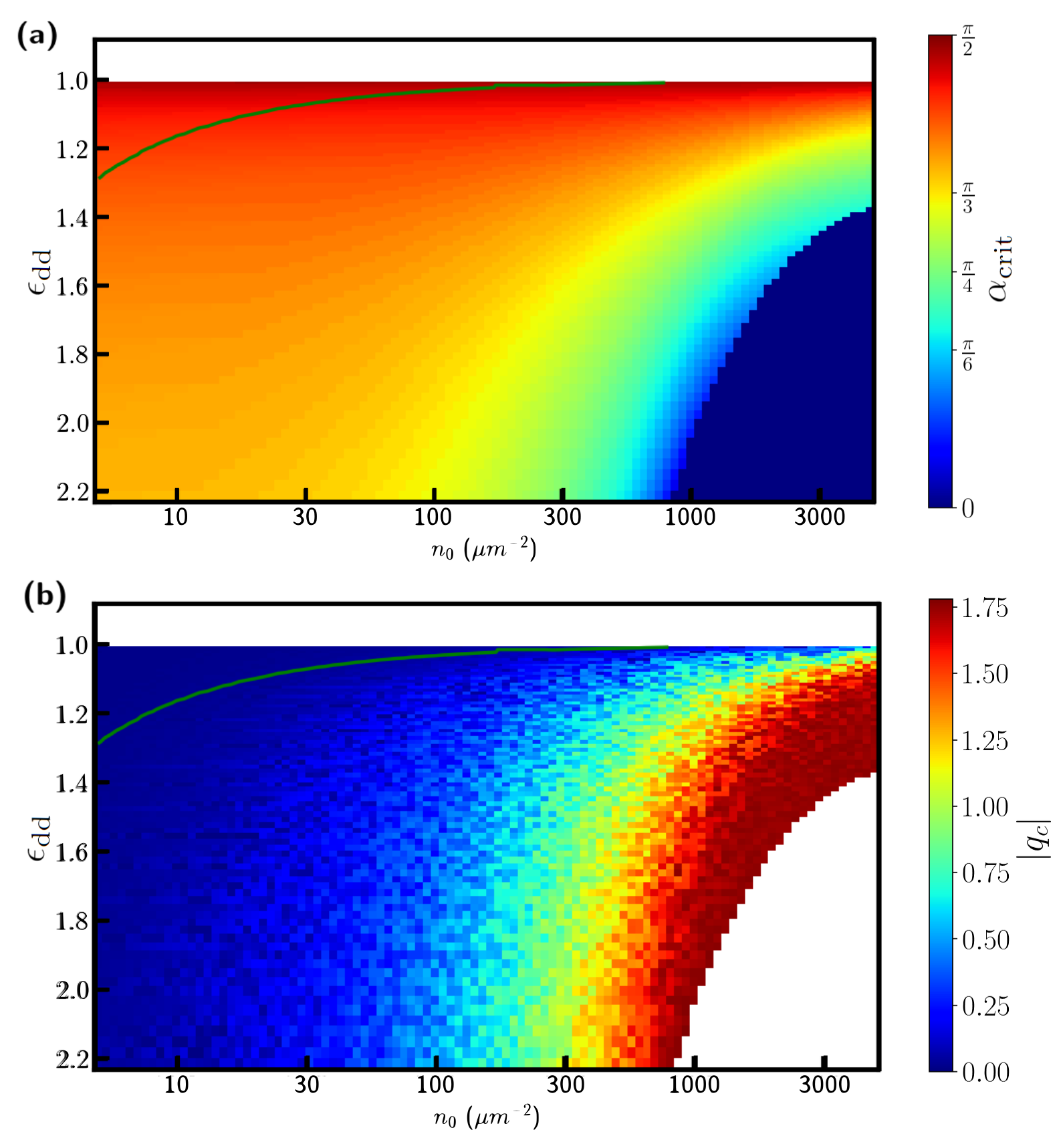}
    \caption{\label{Fig2} \textbf{Identification of instability}.  (a) Values of $\alphac$, the smallest angle $\alpha$ at which there exists one unstable momentum $E_1(\vk)=0$, as a function of the interaction parameter $\edd$ and the mean 2D density $n_0$. The white region for $\edd<1$ is the region where no instability occurs. (b) Norm of the unstable momentum $\vq_c$ at $\alpha=\alphac$. The green line shows the boundary of the region where the instability occurs at $\vq_c=0$. The other parameters are fixed to $\nu_z=1\,$kHz and the gas parameters $m=164\,$a.u., $\add=133\,a_0$, while the changes of $\edd$ are made by varying $\as$.}
\end{figure}

In Sec.~\ref{subsec:energy}, we have noted that 
$E_1(\vk)$ may vanish in some parameter regimes, corresponding to the regime of mean-field instability~\cite{Pitaevskii2016bec,Chomaz2022dpa,Parker:2009,Bohn:2009,Fischer2006soq}. In this regime, perturbation theory breaks down as Eq.\,\eqref{eq:nq} exhibits divergencies. The mean-field instability has been widely studied in dipolar gases in particular due to its dependence on the interaction parameters and on the gas geometry~\cite{Chomaz2022dpa,Parker:2009,Bohn:2009,Fischer2006soq}. The instability can either be of global character, driven by the softening of $\vk=0$ excitations; or of local character, driven by the softening of roton-like excitations at finite momenta.  
In this appendix, we analyze the stability condition and determine the regime of validity of our theory (Sec.~\ref{subsec:perturbth}). We identify the instability by the existence of a vector $\vk$ for which $E_1(\vk)$ vanishes. Our results agree with previous predictions, see e.g.~\cite{Pitaevskii2016bec,Chomaz2022dpa,Parker:2009,Bohn:2009,Fischer2006soq}. For a given set of gas and interaction parameters ($\edd$, $\as$, $n_0$, $\ell_z$), the gas is unstable for all dipole orientations with $\alpha>\alphac$, where $\alphac$ is a critical angle depending on the above parameters.

Figure~\ref{Fig2} (a) shows the value of the critical angle $\alphac$ as a function of $\edd$ and the 2D atomic density $n_0$. 
The gas is always stable if $\edd<1$ (white region in Fig.~\ref{Fig2} (a)). Instability can instead occur if $\add>\as$, as, in this case, dipolar attraction can dominate and drive the instability. 
For $\edd \approx 1$, the instability occurs for $\alphac\approx \pi/2$, corresponding to a dipole orientation nearly colinear with the atomic plane, configuration which maximizes the DDI attractive contribution. When $\edd$ increases, $\alphac$ decreases as a weaker attraction is required to make the gas unstable. 
We also observe that $\alphac$ depends on the 2D density and decreases when the 2D density increases. This effect relates to the local character of the instability in our settings, see discussion of Fig.~\ref{Fig2} (b). At the momentum $\vk_c$ driving the instability, the kinetic term $\frac{\hbar^2\vk_c^2}{2m}$ competes with the total interaction, which scales with the density as $2n_0\tilde{U}(\vk_c)$ and is attractive. 
High-density states are thus destabilized by their increased interaction energy and we find that for large density above $10^{15}$atoms/m$^2$ and small s-wave scattering length, the atomic cloud is unstable even at $\alpha=0$ (dark blue region in Fig.~\ref{Fig2}(a)).

In Fig.~\ref{Fig2} (b), we further report on the norm of the most unstable wavevector $\vq_c=\vk_cl_z/\sqrt{2}$, i.e. for which $E_1(\vk_c)=0$ at $\alpha=\alphac$. First, we note that $\vq_c$ always corresponds to $\phi=0$, i.e. is orthogonal to the dipole plane. The value of $q_c$ allows to tell apart regimes of global and local instabilities, as respectively corresponding to $q_c=0$ and $q_c \neq 0$. The two regimes are separated by the green line in Fig.~\ref{Fig2} (a,b), with the left side (low density, small $\edd$) corresponding to globally unstable cases. In the local instability regime, the value of $q_c$ increases with increasing density and $\edd$, taking values up to $1.75$, meaning $k_c\approx 2.5/l_z$.

In dipolar gases, the mean-field instability is known to be overruled by beyond-mean-field effects~\cite{Chomaz2022dpa} and an extended mean-field description including a beyond-mean-field correction to Eq.\,\eqref{eq:GP} can be applied~\cite{Bisset2019,Ripley2023,Baillie2017}. The beyond-mean-field correction term scales as $\as^{5/2}n_0^{3/2}$ that is with a larger power in density than $E_{\rm int}$. This allows for the stabilization of novel ground states of higher density in the mean-field unstable regime~\cite{Chomaz2022dpa}. Beyond-mean-field corrections are crucial to describe the system in this regime. They may also yield correction in the mean-field stable regime close to the instability threshold as the mean-field interaction term tends to vanish. Yet their quantitative description in this regime is still under debate~\cite{Petter2019}. Furthermore, their effects should be smaller in the present uniform case compared with harmonically trapped ones, where the density also tends to increase close to the instability threshold.  Here we do not consider this extension and restrict our analysis to the mean-field stable regime, meaning that, for a set of $n_0$ and $\edd$ values, we only consider angles with $\alpha<\alphac$.

\end{document}